# Itinerant magnetism in Kondo crystal CeB$_6$ as indicated by polarized neutron scattering

V.P. Plakhty[1,*], L.P. Regnault[2], A.V. Goltsev[3], S.V. Gavrilov[1], F. Yakhou[4], J. Flouquet[2], C. Vettier[5], S. Kunii[6]

[1]*Petersburg Nuclear Physics Institute RAS, Gatchina, 188350 St.Petersburg, Russia*
[2]*CEA-Grenoble, DRFMC/SPSMS/MDN, 17 rue de Martyrs, 38042 Grenoble cedex 9, France*
[3]*Ioffe Physical-Technical Institute RAS, 194021 St. Petersburg, Russia*
[4]*ESRF, rue Jules Horowitz, BP 220, 38043 Grenoble cedex, France*
[5]*Institut Laue-Langevin, 6 Rue J. Horowitz, 38042 Grenoble cedex, France*
[6]*Physics Department, Tohoku University, Sendai 98-77, Japan*

A magnetic Bragg reflection corresponding to the wave vector $\mathbf{k}_{13} = (2\pi/a)[1/2,1/2,1/2]$ of the antiferro-quadrupolar ordering is found in CeB$_6$ in zero magnetic field below the Néel temperature $T_N$. Its intensity is two orders of magnitude weaker than those due to the basic magnetic structure [O. Zaharko *et al.*, Phys. Rev. B **68**, 214401 (2003)]. The peak has a width of the other Bragg reflections below $T_N$, but widens abruptly at $T = T_N$ with simultaneous increase of intensity. Correlation length just above $T_N$ is of the order of 70 Å. The peak intensity decreases to zero at $T \approx 7$ K with no visible anomaly at the antiferro-quadrupolar ordering temperature $T_Q = 3.3$ K. The features of this magnetic ordering are typical for the itinerant magnetism with 5$d$ electron of Ce$^{3+}$ [Yu.S. Grushko *et al.*, phys. stat. sol. (b) **128**, 591 (1985)] being involved.

The studies of the heavy-fermion Kondo-lattice system, CeB$_6$ had started more than 30 years ago,[1] but its highly unusual properties are being still extensively debated. Cerium hexaboride crystallizes in the CaB$_6$ – structure (space group *Pm*3*m*) with a B$_6$ molecule in the form of a regular octahedron and the Ce ion situated, respectively, at the corner and at the center of the simple cubic unit cell with $a = 4.141$ Å).[1] A theoretical study of CaB$_6$ – type materials[2] indicates that each B$_6$ octahedron requires 2 electrons from Ce to stabilize a 3D boron network. Two electrons are left on Ce$^{3+}$, 4$f$ and 5$d$.[3] While 4$f$ system is studied in detail, magnetic interactions among 5$d$ electrons have never been observed. We present in this communication indications of 5$d$–electrons itinerant magnetism.

It has been established that the crystalline electric field of the cubic boron environment splits the Ce$^{3+}$ multiplet 4$f^1$ into a ground-state quartet $\Gamma_8$ and a doublet $\Gamma_7$ with a gap of 47 meV.[4] At the temperature range of interest ($T < \sim 7$ K) one can neglect admixture of excited level. The resistivity grows logarithmically, attaining its maximum at $T \approx 3.2$ K.[5] The Kondo temperature was estimated as $T_K \approx 8$ K,[6] but then it was revised down to $T_K \approx 1$ K on the basis of magnetic susceptibility data.[7] The phase diagram in this region is very unusual. Two specific heat anomalies were observed at about 3.3 K and 2.4 K in the absence of external magnetic field.[8] The results of a zero-field neutron diffraction experiment,[9] were explained by antiferromagnetic ordering with wave vectors $\mathbf{k}_4 = (2\pi/a)[m, m, 0]$ and $\mathbf{k}_5 = (2\pi/a)[m, m, 1/2]$ ($m = 1/4$) below the second maximum that was attributed to the Néel temperature $T_N = 2.4$ K. (The wave vectors are numbered according to Kovalev.[10]) No superstructure reflections were detected in between two maxima of specific heat. Later on, a series of superstructure reflections with the wave vector $\mathbf{k}_{13} = (2\pi/a)[1/2, 1/2, 1/2]$ was found by neutron diffraction in this temperature region in the presence of an external magnetic field.[11]. This observation was explained by an ordering of the quadrupolar moments $Q$ and $-Q$ at $T_Q = 3.3$ K, which would split the ground-state quartet $\Gamma_8$ into two doublets. Due to the quadrupolar ordering, the external magnetic field would induce antiferromagnetic spin arrangement with the same wave vector $\mathbf{k}_{13}$. All these results indicated the existence of three phases: the non-ordered paramagnetic phase I ($T_Q < T$), the antiferro-quadrupolar (AFQ) phase II ($T_N < T < T_Q$) and the antiferromagnetic phase III ($T < T_N$). At that point the CeB$_6$ saga seemed to be over. However, further experimental and theoretical works led to a renewed interest in CeB$_6$.

Antiferromagnetic ordering in the phase II under magnetic field has also been observed by $^{11}$B-NMR,[12,13] but with a different structure described by 3 arms of the wave-vector star $\{\mathbf{k}_{10}\} = (2\pi/a)[0, 0, 1/2]$. The $\mu$SR results[14,15] have been in contradiction with the magnetic structures suggested by both the neutron diffraction[9,11] and the NMR[12,13] studies. In order to explain consistently all experimental data on the phase II field–induced magnetic structure, the octupolar moments $T_{xyz}$ have been considered.[16] It has been shown that the AFQ ordering in phase II induces dipoles and octupoles in presence of magnetic field, with the octupoles playing a crucial role.[17] Recently, the antiferromagnetic structure in magnetically ordered phase III has been reexamined.[18] It has been shown that a new multi–$\mathbf{k}$ modulated model is in a better agreement with neutron diffraction and zero-field polarization analysis. Moreover, it is consistent with the other data[13-15] as well. The translational symmetry of this structure is described by the wave vectors $2\mathbf{k}_4 - \mathbf{k}_5$ and $2\mathbf{k}_4' - \mathbf{k}_5'$, where $\mathbf{k}_4' = (2\pi/a)[m, -m, 0]$, $\mathbf{k}_5' = (2\pi/a)[m, -m, 1/2]$, and $m = 1/4$. (A very weak intensity was also detected at the (1/2,1/2,1/2) position, but it had no polarization dependence, and was explained by the second order contamination.) The magnetic moments of the nearest $Ce^{3+}$ ions in a (1,0,0) plane are aligned along [1,1,0] and [1,–1,0]. According to this model, a moment modulation occurs within one plane with the given above wave vectors $\mathbf{k}_4$ and $\mathbf{k}_4'$, as well as a strong difference of the moment values at $z = 0$ and $z = 1$. This very complicated magnetic structure is explained by competition of the AFQ order with the dipolar and octupolar order. In spite of the consistency achieved, Kasuya in a series of publications[19,20] has claimed that there is no AFQ ordering in phase II, but observed intensities arise from antiparallel displacements of Ce along a $\langle 0,0,1 \rangle$ direction, which are ordered with the wave vector $\mathbf{k}_{13}$. To distinguish two types of ordering we have undertaken synchrotron X-ray diffraction experiment.[21] A series of very weak reflections from the planes $\{h/2, k/2, l/2\}$ have been observed below $T_Q$. However, because of a low X-ray energy of 5.2–5.7 keV and the vicinity to the Ce $L_{III}$ absorption edge we have failed to make proper corrections that would allow to distinguished between the Kasuya model and the ordering of quadrupoles with the form factor.[22] Later on, it has been shown[23] that taking into account the Thomson scattering makes the data[21] consistent with the AFQ order. Reflections corresponding to $\mathbf{k}_{13}$ were also detected in X-ray resonance scattering experiment,[24] but no analysis of their actual nature was made.

In a further to attempt to resolve the issue, we have carried out the present neutron polarization analysis experiment in a low magnetic field to search for possible displacements of B atoms, which could explain the $\mathbf{k}_{13}$ superstructure. The experiment has been performed on the single crystals used by Effantin *et al.*[11] A mosaic of 5 crystal plates has been prepared by orientation of every one and fixing on aluminum plate without any glue to avoid strong incoherent scattering by hydrogen. The vertical axis has been chosen to be [1,–1,0], with the [1,1,0] axis being horizontal and perpendicular to the plate. The [0,0,1] axis belongs to the scattering plane as well as to the crystal plate. The plate with the crystal mosaic has been mounted in an orange cryostat installed on the polarized three-axis spectrometer IN22 of thermal neutrons at the ILL high-flux reactor. The neutrons with $k_i = 2.62$ Å$^{-1}$ have been used with a PG filter to suppress the second order harmonic in the incident beam. The filter has decreased the $2k_I$ contamination to the level less than $10^{-4}$ in comparison to the main component. The total mosaic disorientation is about 1.5° as seen from the rocking curve of nuclear reflection (0,0,2) in Fig. 1. This reflection has been measured by means of $w$-scan with the neutron polarization $\mathbf{P}_0$ along the momentum transfer $\mathbf{Q}_{hkl}$. The guide field maintaining the neutron polarization was about 10 Oe. It should be kept in mind that the initial neutron polarization changes its sign (flips) due to the spin components orthogonal to both $\mathbf{Q}$ and $\mathbf{P}_0$.[25] In the geometry used for the experiment, the nuclear scattering is completely non-spin-flip while any magnetic scattering flips neutron spin. The rocking curve for the nuclear reflection (0,0,2) (Fig. 1) has been measured in both modes, spin-flip (SF) and non-spin-flip (NSF). Flipping ratio $I^{NSF}/I^{SF}$ has been measured to be about 25. Similar non-spin-flip scans with much longer measuring time have been made through the positions of reflections (1/2,1/2,1/2), (1/2,1/2,3/2), (3/2,3/2,1/2), (1/2,1/2,5/2), (3/2,3/2,1/2), corresponding to $\mathbf{k}_{13}$ wave vector. No NSF signal could be detected within noise, with upper limits to the ratios $I(h/2,k/2,l/2):I(0,0,2) \sim 5 \cdot 10^{-6}$. Using Eq. (2) from Ref. 21, the upper limits for the combined atomic displacements[19,20] can be estimated as $\Delta_B < 10^{-4}$ Å, $\Delta_{Ce} < 10^{-3}$ Å, or $\Delta_B < 8 \cdot 10^{-4}$ Å, $\Delta_{Ce} < 2 \cdot 10^{-3}$ Å. Actually these limits should be even less due to inevitably high extinction for the (0,0,2) reflection. (Remind that $\Delta_{Ce} < 10^{-4}$ Å from the X-ray data.[21]) Therefore, no significant displacements of either Ce or B have been detected.

As shown in Fig. 1, a SF signal has been detected at the position (1/2,1/2,1/2) in zero field, indicating that this scattering is magnetic in origin. The $w$–profile of reflection (0,0,2) clearly shows 5 peaks corresponding to the number of crystals mounted into mosaic. The profile of reflection (1/4,1/4,1/2) due to basic magnetic structure as well as the (1/2,1/2,1/2) profile are fitted by 5 Gaussians with the constraints for the peak widths and the distances between the peaks taken from the (0,0,2) fit. (The relative intensities of the 5 different contributions are not necessarily the same for all the reflections, since the 5 crystals might be tilted with different orientations.) It is seen that the total width of reflection (1/2,1/2,1/2) is the same as those of (0,0,2) and (1/4,1/4,1/2). The integrated intensity ratios are $I(0,0,2) : I(1/4,1/4,1/2 : I(1/2,1/2,1/2) = 1 : 3.13(6) \cdot 10^{-2} : 2.1(4) \cdot 10^{-4}$. The magnetic reflection (1/2,1/2,1/2) is extremely weak, and therefore the corresponding component of the magnetic moment leading to the scattered intensity is about one order of magnitude smaller than the average value $\langle m \rangle \approx 0.35$ $m_B$ for the basic magnetic structure.[18]

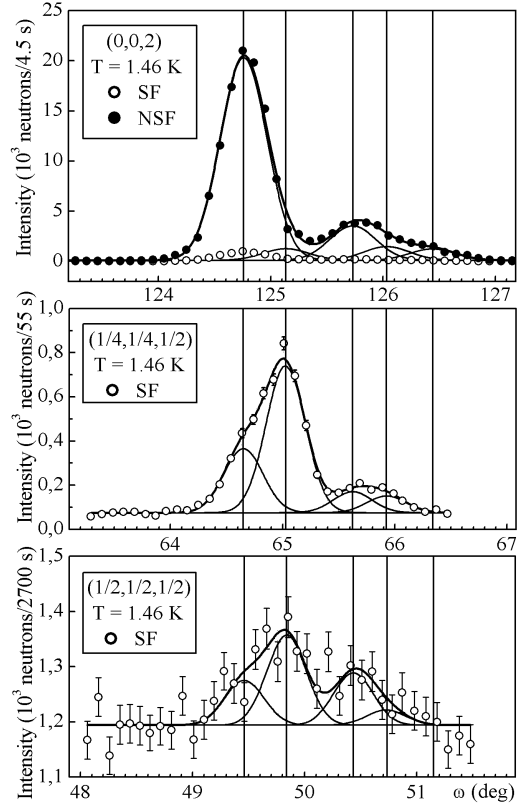

FIG. 1. Rocking curves obtained with the nuclear reflection (0,0,2) and with magnetic reflections (1/4,1/4,1/2), (1/2,1/2,1/2). The scans are shifted to make the peak positions coinciding.

Longitudinal scans (along **Q**) through the maximum of reflection (1/2,1/2,1/2) are shown in Fig. 2 for $T = 1.46$ K (a) and $T = 2.47$ K (b), (c). At 1.46 K, the profile is fitted by that of the nuclear Bragg reflection (1,1,1) measured with the wave length $l/2$ without the PG filter. At 2.47 K the peak becomes much broader. The best fit by a Lorentzian results in its position $h_m = 0.476(7)$, intensity in the maximum $I_m = 210(41)$ and the full width at half-maximum $2\Gamma = 0.069(31)$. The Lorentzian width corresponds to the correlation length $x \sim 70$ Å. (The maximum is shifted from $h = 0.5$. If this shift is due to an incommensurate structure, then there should be a peak on the other side of $h = 0.5$ as shown in Fig. 2(c). However, this shift is most probably an artifact that could arise due to the focusing geometry. Since the shift is in the limits of 3.4 standard deviations, we assume that the peak remains commensurate at the wave vector $\mathbf{k}_{13} = (2p/a)[1/2,1/2,1/2]$.

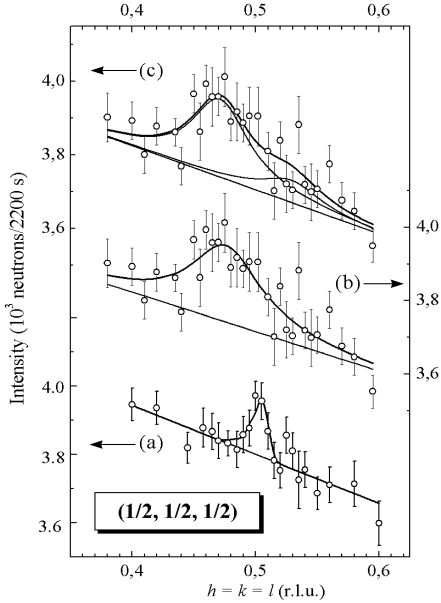

FIG. 2. Spin-flip longitudinal scans through the maximum of reflection (1/2,1/2,1/2). (a) $T = 1.46$ K, (b) and (c) $T = 2.47$ K, with two different fits as explained in the text.

Temperature dependence of intensity at the (1/2,1/2,1/2) position is shown in Fig. 3. The intensity develops below about 7 K, i.e., at a temperature three times higher than the Néel point. No anomaly could be detected at $T_Q$. Below $T_N$, the peak intensity falls abruptly by a factor of two, while the peak width decreases to reach that of a Bragg reflection as shown in Fig. 2.

Polarization analysis has been carried out at $T = 1.46$ K <

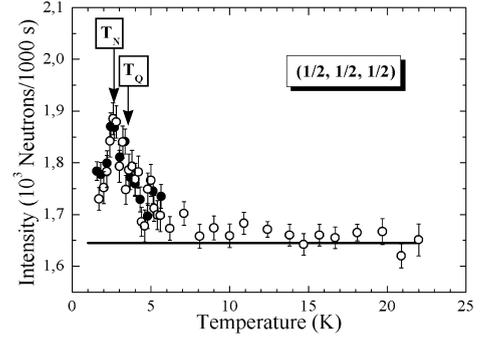

FIG. 3. Spin-flip intensity temperature dependence at the maximum of (1/2,1/2,1/2) reflection. Open and filled circles present two sets of measurements. Straight line shows the background level.

$T_N$ and $T = 2.47$ K $> T_N$ at the maxima of the reflections (1/4,1/4,1/2) and (1/2,1/2,1/2) with different orientations of $\mathbf{P}_0$: $\mathbf{P}_x$ – along **Q**, $\mathbf{P}_y$ – orthogonal to $\mathbf{P}_x$ in the scattering plane, and $\mathbf{P}_z$ – orthogonal to the scattering **xy** plane. The spin-flip intensities are collected in Table I. The

TABLE I. Spin-flip intensities $I_x^{SF}$, $I_y^{SF}$, $I_z^{SF}$ of reflections (1/4,1/4,1/2) and (1/2,1/2,1/2) with polarization $\mathbf{P}_x$, $\mathbf{P}_y$, $\mathbf{P}_z$ along **x**, **y** and **z** axes, respectively, as explained in the text.

|  | (1/4,1/4,1/2) | | (1/2,1/2,1/2) | |
|---|---|---|---|---|
| $T$(K) | 1.46 | 2.47 | 1.46 | 2.47 |
| $\mathbf{P}_x$ | 25836(376) | 2236(216) | 216(47) | 307(63) |
| $\mathbf{P}_y$ | 25320(372) | 1892(212) | 80(46) | 83(61) |
| $\mathbf{P}_z$ | 2109(212) | 1076(204) | 71(46) | 90(61) |

intensities of reflection (1/4,1/4,1/2) at $T < T_N$ are in agreement with those reported in Ref. 18, where, as follows from the results, a single domain crystal has been investigated, and the associated spin direction is found to be [1,−1,0]. The finite value of $I_z^{SF}$ in our case can be explained by small amount of the other domains, which does not influences the conclusion on the spin direction along the $\langle 1,-1,0 \rangle$-type axes. The (1/4,1/4,1/2) intensity $I_x^{SF}$ falls by a factor of 12(1) for temperatures above $T_N$. The relative components for different polarizations changes are modified when crossing $T_N$, indicating that above $T_N$ there is an admixture of critical scattering with the intensity ratio similar to that observed at $T < T_N$ and paramagnetic scattering with $I_x^{SF} : I_y^{SF} : I_z^{SF} = 2:1:1$.[25] The experimental uncertainties for the polarized intensities of reflection (1/2,1/2,1/2) are too large to draw any conclusion about the moments direction. Nevertheless, the behaviors of reflections (1/2,1/2,1/2) and (1/4,1/4,1/2) are clearly different. First, the broad peak at $T > T_N$ does not arise from critical scattering, which would be down by about one order of magnitude, as it is observed for the (1/4,1/4,1/2) reflection. Second, the polarized intensities below $T_N$ do not differ drastically to those observed above $T_N$, for $\mathbf{P}_x$ [0.7(2)],

$P_y$ [1.0(9)] and $P_z$ [0.8(7)], in sharp contrast to what is observed in the case of reflection (1/4,1/4,1/2) [12(1), 13(2) and 2.0(4), respectively]. This indicates that the spin direction for the long-range order associated with $k_{13}$ at $T < T_N$ and for the fluctuations at $T > T_N$ is probably the same.

The behavior of reflection (1/2,1/2,1/2) indicates an origin of magnetic ordering with $k_{13}$, which is not the same as the basic multi-k magnetic structure of localized 4f Ce spins. This behavior is typical for the interacting electrons in the conduction band, which is filled in $CeB_6$ by hybridized 5d electrons of cerium$^3$ and 2p electrons of boron according to the band calculations.$^{26,27}$ At definite conditions, these interaction can be stronger than those between the moments of localized 4f electrons, and as a result the antiferromagnetic fluctuations of itinerant electrons may appear at $T \gg T_N$ (Remind that in our case the antiferromagnetic fluctuations arise at $T \approx 3T_N$). As shown in Ref. 28, where the conduction states are described by three nearly spherical pockets centered at the X points of the simple cubic Brillouin zone, the Fourier transform $J(\mathbf{q})$ of RKKY interaction between localized 4f electrons mediated by hybridized 5d(2p) conduction electrons has a maximum at $\mathbf{q} = \mathbf{k}_{13}$ (R point of the Brillouin zone) in addition to the maximum at $\mathbf{k}_5 = (2\pi/a)[1/4,1/4,1/2]$ that defines the basic magnetic structure. The maximum of $J(\mathbf{q})$ corresponds to the maximum of susceptibility of the conduction electrons, $J(\mathbf{q}) \propto c(\mathbf{q})$. Therefore, antiferromagnetic fluctuations at $\mathbf{q} = \mathbf{k}_{13}$ are likely to develop. It is well known that in a metal with the simple cubic lattice the antiferromagnetic fluctuations with the wave vector $\mathbf{k}_{13}$ may be considerably enhanced by the Stoner factor.$^{29,30}$ Due to interactions between conduction electrons and the localized spins the antiferromagnetic transition at $T_N$ strongly influences the energy spectrum of conduction electrons and, as a result, changes the intensity and the width of the susceptibility peak at $\mathbf{q} = \mathbf{k}_{13}$. Taking into account the translation relation $\mathbf{k}_{13} = \mathbf{k}_4 + \mathbf{k}_5$, one can assume that these interactions and interactions between multipolar moments of localized 4f electrons should induce the long-range ordering of the fluctuations with $\mathbf{k}_{13}$ at $T_N$.

When summarizing the present results together with those of Ref. 21 we can conclude that neither Ce nor B displacements are detected down to the level $\sim 10^{-4}$ Å, which rules out the model of pair displacements.$^{19,20}$ However, magnetic diffuse scattering is observed at the reciprocal lattice point (1/2,1/2,1/2) below $T \approx 7$ K, three times higher than $T_N$. The fluctuations build up with no visible anomaly at $T_Q = 3.3$ K, which means that the quadrupolar ordering does not strongly influence the energy spectrum of conduction electrons. The highest correlation length is about 70 Å just above the Néel point. Below $T_N$, the peak of diffuse scattering transforms into narrow Bragg reflection, with the peak intensity being abruptly decreased. An additional long-range magnetic order appears with the wave vector $\mathbf{k}_{13} = (2\mathbf{p}/a)[1/2,1/2,1/2]$ and with the moment value about $10^{-1}$ of that for the basic magnetic structure.$^{18}$ All these features are typical for the itinerant magnetism, with 5d electrons of $Ce^{3+}$ being involved.

We are grateful to S.V. Maleyev and V.P. Mineev for the fruitful discussions. Two of us (V.P. Plakhty and S.V. Gavrilov) acknowledge a partial financial support from the State Program on quantum macrophysics (Grant 10002-25/П-03/040-58/100603), from the Russian Foundation for Fundamental Researches (Project № 02-02-16981), from the Grant SS-1671.20032.